\documentclass[aps,pra,10pt,twocolumn,amsmath,amssymb]{revtex4-2}

\usepackage[OT1]{fontenc}
\usepackage[utf8]{inputenc}

\usepackage{mathptmx}
\DeclareMathAlphabet{\mathcal}{OMS}{cmsy}{m}{n}

\usepackage{bm,bbold}
\usepackage{dsfont}
\usepackage[colorlinks=true,citecolor=blue,breaklinks=true,linkcolor=blue,linktocpage=true,urlcolor=blue,pagebackref=false]{hyperref}
\usepackage{orcidlink}
\usepackage{graphicx}

\newcommand{\sect}[1]{\par\textit{#1}.---\ignorespaces}
\newcommand{\llangle}{\langle\!\langle}
\newcommand{\rrangle}{\rangle\!\rangle}
\newcommand{\gammaAA}{\gamma_\text{AA}}
\newcommand{\gammaFB}{\gamma_\text{FB}}

\newcommand{\void}[1]{}

\begin{document}

\title{Measurement protocol for non-adiabatic geometric phases of Floquet states}

\author{Carlos Martín-Fernández\,\orcidlink{0009-0008-7910-9610}}
\author{Gloria Platero\,\orcidlink{0000-0001-8610-0675}}
\author{Sigmund Kohler\,\orcidlink{0000-0003-3668-8030}}

\affiliation{Quantum Advanced Research Center (QuARC), CSIC, 28049 Madrid, Spain}
\affiliation{Instituto de Ciencia de Materiales de Madrid (ICMM), CSIC, 28049 Madrid, Spain}

\date{\today}

\begin{abstract}
Periodically time-dependent quantum systems have Floquet solutions.  They
consist of a phase factor and a Floquet mode which shares the periodicity
of the driving.  Consequently, any such solution possesses a non-adiabatic
geometric phase known as the Aharonov-Anandan (AA) phase.  It is
gauge-invariant and, therefore, is experimentally accessible.
We propose a protocol for the measurement of AA phases based on the adiabatic
change of the phase of the non-adiabatic driving.  We argue that the
experimental requirements of the protocol are those of previous
measurements of adiabatic phases.
\end{abstract}

\maketitle

\sect{Introduction}
The phase acquired by a quantum system under cyclic adiabatic parameter
variation consists of a dynamical and a geometric part.
The latter depends only on the trajectory in parameter space and is gauge invariant
\cite{BerryPRSA84b}.
Geometric phases are not only of interest from a fundamental perspective,
but are also relevant for practical calculations of electronic properties
of Bloch bands \cite{RestaJP00,XiaoRMP10}.  They play a role for
holonomic quantum computation \cite{ZanardiPLA99,PachosPRA99} and for qubit
control by shortcuts to adiabaticity \cite{GueryOdelinRMP19}.  In the
context of solid-state quantum information processing, direct
measurement of Berry phases has been proposed \cite{FalciNL00} and
experimentally realized with superconducting qubits \cite{LeekS07,
ZhangPRA17}, nitrogen-vacancy centers \cite{MaclaurinPRL12},
and quantum dots \cite{ZhouNC25}.

The non-adiabatic generalization of the Berry phase, the Aharonov-Anandan
(AA) phase, emerges when a state vector undergoes a cyclic evolution in
Hilbert space \cite{AharonovPRL87}.  Corresponding experiments have been
performed with photonic systems \cite{WangOL16}, Bose-Einstein condensates
\cite{DoPRR19}, nitrogen-vacancy centers \cite{WoodPRB20}, and nuclear
spins \cite{BengsJCP23}.  These experiments have in common that the systems
are not explicitly time dependent, except for the application of pulses that
compensate a dynamical phase.

The AA phase also plays a role in Floquet systems, which possess solutions
with Floquet structure, i.e., they consist of
a phase factor and a time-periodic Floquet mode.  Various phases are associated with
these modes. First, there is the total phase acquired by a Floquet solution
during one driving period. It is determined by the quasienergies
\cite{ShirleyPR65,SambePRA73}, which can be separated into a dynamical contribution
stemming from the time-averaged energy expectation value and an AA phase
\cite{MooreJPA90b}.  When, in addition, the driving parameters undergo a
cyclic adiabatic variation, a Floquet-Berry (FB) phase emerges
\cite{WeinbergPR17}.

This raises the question of whether AA phases can be measured directly.  In
this Letter, we present a corresponding protocol which is based on two
findings derived below.  First, under slow variation of the
driving phase from $0$ to $2\pi$, a Floquet state acquires an adiabatic FB
phase that matches the non-adiabatic AA phase.  Second, a protocol similar
to that of Ref.~\cite{LeekS07} can be employed for the measurement of the
FB phase of qubits under fast and strong driving.
Moreover, we explore the conditions for experimental realizations.

\sect{Floquet states under adiabatic phase control}
We consider a periodically time-dependent Hamiltonian $H(t) = H(t+T)$ with driving frequency
$\Omega = 2\pi/T$. By means of the Floquet theorem,
any solution of the corresponding Schrödinger equation can be decomposed into terms of the form
$|\psi(t)\rangle=e^{-iq t}|\phi(t)\rangle$ with a phase factor determined by the quasienergy $q$
and a Floquet mode $|\phi(t)\rangle=|\phi(t+T)\rangle$.  Both can be computed using the Floquet
equation $\mathcal{H}(t)|\phi(t)\rangle = q|\phi(t)\rangle$ with
$\mathcal{H}(t) = H(t)-i\partial_t$, which is an eigenvalue equation in Sambe space, i.e.,
Hilbert space tensored with the space of $T$-periodic functions \cite{ShirleyPR65,SambePRA73}.
The $T$-periodicity of the Floquet modes gives rise to a non-adiabatic geometric phase, namely
the AA phase \cite{AnandanPLA88, MooreJPA90b}
\begin{equation}
    \gammaAA= \int_0^T dt\, \langle \phi(t)| i\partial_t |\phi(t)\rangle.
    \label{anandan_formula}
\end{equation}
It plays a prominent role in Floquet theory due to the relation between the time-averaged
energy expectation value and the quasienergy, $\bar E = q + \gammaAA/T$ \cite{MooreJPA90b}.
The geometric nature of the AA phase indicates that it is measurable.

To derive a protocol that provides direct access to $\gammaAA$, we consider the phase
shifted Hamiltonian $H(t+\theta/\Omega)$, where the phase $\theta$ may undergo adiabatic variation.
Obviously, a constant phase shift turns the Floquet mode into $|\phi(t+\theta/\Omega)\rangle$.
By contrast, it does not affect the quasienergies or any time-averaged expectation value.
In particular, the mean energies and the AA phases remain the same.

For sufficiently slow variation of the driving parameters, adiabatic following of the wavefunction
to the Floquet state is expected \cite{WeinbergPR17}. This is associated with a FB phase,
which formally is like the usual Berry phase, but with the adiabatic eigenstates of a Hamiltonian and
the Hilbert space structure replaced by the Floquet modes
and the inner product of Sambe space, respectively \cite{WeinbergPR17}.
For $|\phi(t+\theta/\Omega)\rangle$ the derivatives with respect to $t$ and $\theta$ are essentially the same.  Therefore, we expect that a cyclic adiabatic variation of $\theta$ leads to a FB phase that is related to the AA phase.

For a more rigorous analysis, we insert the adiabatic-following ansatz
$|\psi(t)\rangle = e^{i\varphi(t)}|\phi(t+\theta/\Omega)\rangle$ into the
Schr\"odinger equation to obtain for the total phase the equation of motion
\begin{equation}
\dot\varphi = -q
+\langle\phi(t+\theta/\Omega)|\,i\partial_\theta\,|\phi(t+\theta/\Omega)\rangle
\dot\theta,
\label{eq:phidot}
\end{equation}
where we have used the Floquet equation to obtain the term $-q$.
The phase acquired during $N$ driving periods follows by time integration from $0$ to
$NT$ and reads $\varphi = -qNT + \gammaFB$ with the FB phase following from the last term in Eq.~\eqref{eq:phidot}.
To evaluate it for slow variation of $\theta$, we assume that in any time interval $[(n-1)T,nT]$, the time-dependent
$\theta(t)$ can be replaced by the constant value $\theta_n = \theta(nT)$.
This allows us to write the integral as a sum of $N$ integrals over one driving period, such that
\begin{align}
\gammaFB
={}& \sum_{n=1}^{N}\frac{\theta_n-\theta_{n-1}}{\Omega T}
\int_{(n-1)T}^{nT} \!\! dt\langle\phi(t+\tau_n)|
i\partial_t |\phi(t+\tau_n)\rangle
\nonumber \\
={}& \gammaAA ,
\label{eq:equivalence}
\end{align}
with the shorthand notation $\tau_n = \theta_n/\Omega$.  To obtain the last equation, we have used the $T$-periodicity of the Floquet modes, which implies that the integrals do not depend on $\tau_n$. Therefore, they are all equal to $\gammaAA$, see Eq.~\eqref{anandan_formula}.  The remaining telescoping sum becomes $\theta_N-\theta_0$, which for $\theta$-variation from $0$ to $2\pi$ cancels the denominator. For a detailed derivation, see the \cite{supplement}.
The practical use of Eq.~\eqref{eq:equivalence} is to link the (non-adiabatic) AA phase and the FB phase resulting from adiabatic $\theta$-variation. This relation forms the basis of the protocol derived below.

Equation \eqref{eq:equivalence} has been derived for an arbitrary adiabatic variation of the drive phase. For the special case of a variation linear in time, an elegant shortcut to this result is enabled by the known relations between mean energies and quasienergies, $\bar E = q+\gammaAA/T$ \cite{MooreJPA90b} and $\bar E = q-\Omega(\partial q/\partial\Omega)$ \cite{FainshteinJPB78}, hence $\partial q/\partial\Omega = -\gammaAA/2\pi$.
For $\theta = \Omega t/N$ with $N\gg 1$, the effective drive frequency is $\Omega'=\Omega+\Omega/N$, such that, to first order in $1/N$, the quasienergy becomes $q(\Omega') = q(\Omega) + (\partial q/\partial\Omega)\Omega/N$.  Then for the evolution over time $t=NT$, a Floquet state acquires the additional phase $-[q(\Omega')-q(\Omega)] NT = \gammaAA$.
A non-trivial aspect is that this additional phase depends only on the geometric phase $\gammaAA$ and not on any dynamical phase.

\begin{figure}
    \centerline{\includegraphics[width = 0.475\textwidth]{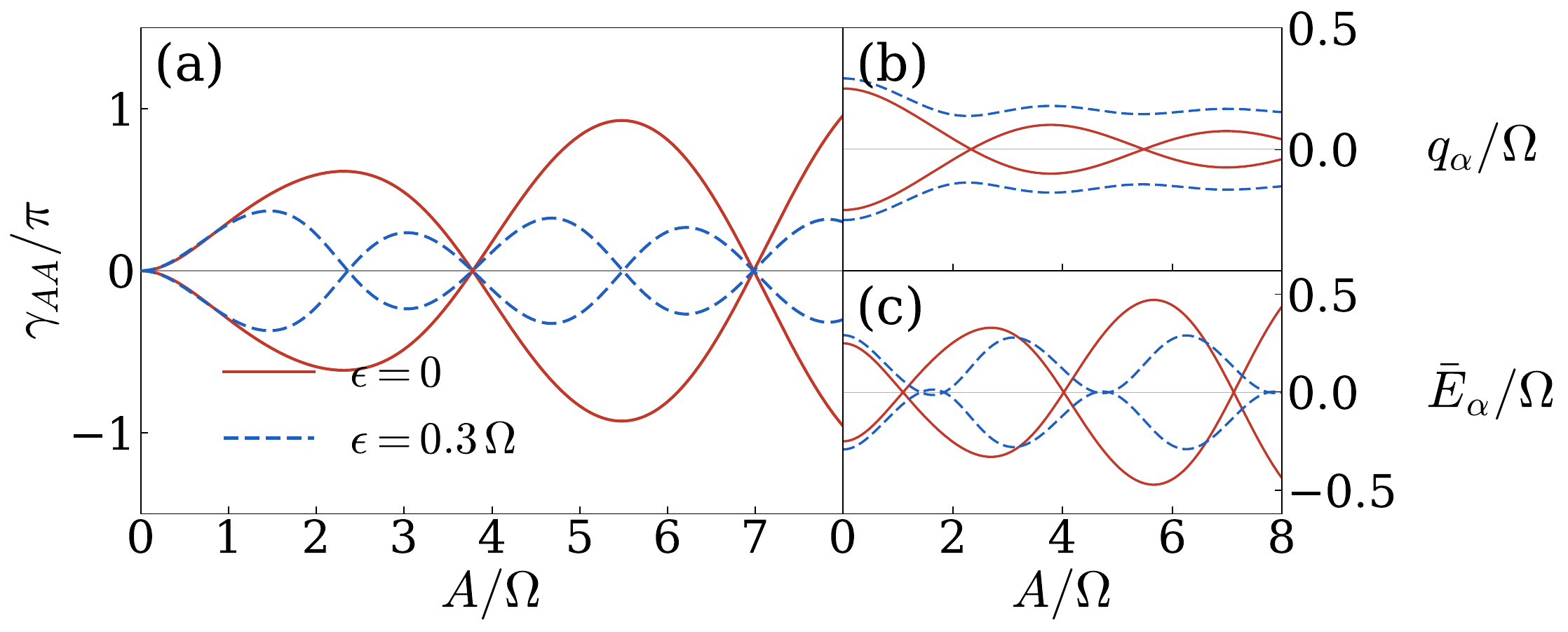}}
    \caption{Floquet spectrum for detunings $\epsilon=0$ (solid red) and $\epsilon=0.3\,\Omega$ (dashed blue) and tunneling $\Delta=\Omega/2$.
    The panels show the AA phases (a), the quasienergies (b), and the mean energies (c).}
    \label{fig:quasi}
\end{figure}

\sect{Model Hamiltonian}
As a paradigmatic model, we employ the driven two-level system (TLS)
\begin{equation}
    H(t)=\frac{\Delta}{2}\sigma_x+\frac{1}{2} \big(\epsilon + A\cos(\Omega t) \big)\sigma_z,
    \label{Hamiltonian}
\end{equation}
with tunnel splitting $\Delta$, detuning $\epsilon$, and amplitude $A$, while $\sigma_{x,z}$ denote
the Pauli matrices.  Its Floquet spectrum is shown in Fig.~\ref{fig:quasi}, where we have chosen
a symmetric Brillouin zone such that $-\Omega/2 < q_\pm \leq \Omega/2$.  In the absence of detuning,
the quasienergies are approximately given by a Bessel function \cite{GrossmannEL92} and exhibit exact
crossings \cite{PeresPRL91,KohlerQ26}. For detuning $0.3\Omega$, the quasienergies depend only weakly on the amplitude.
Nevertheless, the corresponding AA phases shown in Fig.~\ref{fig:quasi}(a) vary significantly over a range of size $\pi$.
An intriguing detail is that $\gammaAA$ as a function of $A/\Omega$ has zeros that are practically independent of the detuning. Moreover, for $\epsilon>0$ additional zeros emerge.  In the high-frequency limit $\Omega\gg\Delta$, this characteristic is exact \cite{supplement}.
Our goal is to derive a protocol under which the system returns to its initial state with a
phase factor that is given solely by $\gammaAA$.

A relevant symmetry of the Floquet Hamiltonian for $H(t)$ in
Eq.~\eqref{Hamiltonian} is the chirality $S = \sigma_y\otimes (t\to -t)$,
with $S\mathcal{H}S^{-1} = -\mathcal{H}$ which pairs Floquet modes with
opposite quasienergy.  Since $S^2 = \mathbb{1}$,
$S|\phi_\pm(t)\rangle = \sigma_y|\phi_\pm(-t)\rangle =
|\phi_\mp(t)\rangle$ with opposite quasienergy $q_-=-q_+$. This leads to the
reflection symmetry of the Floquet spectrum at the horizontal axis visible in Fig.~\ref{fig:quasi}.

\sect{Compensating the dynamical phase}
The total phase acquired during adiabatic following also contains a
dynamical phase, which usually dominates.  For a Floquet
problem it is determined by the quasienergy \cite{WeinbergPR17} appearing
in Eq.~\eqref{eq:phidot}.  To cancel the dynamical phase, we adapt the
spin-echo idea of Ref.~\cite{LeekS07} to our needs.  There, after a first
cycle, a $\pi$-pulse was applied to transfer the adiabatic eigenstate to its
partner with opposite adiabatic energy.  Then the cycle was traversed a
second time in the inverse direction. As a result, the dynamical phases cancel each
other, while the geometric phases add up.  The formal requirement for this
is a chirality which pairs the eigenstates with opposite energy.

For the present Floquet system, the chirality $S$ is rather peculiar, as it contains the
time inversion $t\to -t$, which may affect the spin-echo.
While in an experiment direct inversion of $t$ seems impossible, the $T$-periodicity
of the Floquet modes provides a solution. It implies $|\phi(-kT)\rangle = |\phi(kT)\rangle$
for any integer $k$, such that at times $t=kT$, the mapping $t\to -t$ does not
affect the Floquet modes.  Therefore, at stroboscopic times, the spatio-temporal
chirality $S$ of the Floquet Hamiltonian $\mathcal{H}$ acts in the same way as the
chirality $\sigma_y$ of the Hamiltonian $H(t)$.  This requires that the spin-echo pulse
is applied at a multiple of the driving period, i.e., that the protocol time must be a
multiple of the driving period, a condition that we have already assumed in
the derivation of the fundamental relation \eqref{eq:equivalence}.
It must also be noted that switching to the chirality-related mode not only leads to the opposite
dynamical phase but also inverts the FB phase.  The latter sign change can be reversed
by varying the driving phase from $2\pi$ to $0$, as we explicitly
demonstrate in the \cite{supplement}.

A proper protocol must also consider the preparation of an initial Floquet mode, which can
also be achieved by adiabatic following. In the absence of
driving, the system is let to relax to its ground state.  Formally, the ground state is
a Floquet state for $A=0$, because it fulfills the Floquet equation.  Therefore, one can
simply adiabatically ramp up the amplitude until the desired value of $A$ is reached.
This turns the ground state into the corresponding Floquet mode.

\begin{figure}
\centerline{\includegraphics[width=.9\columnwidth]{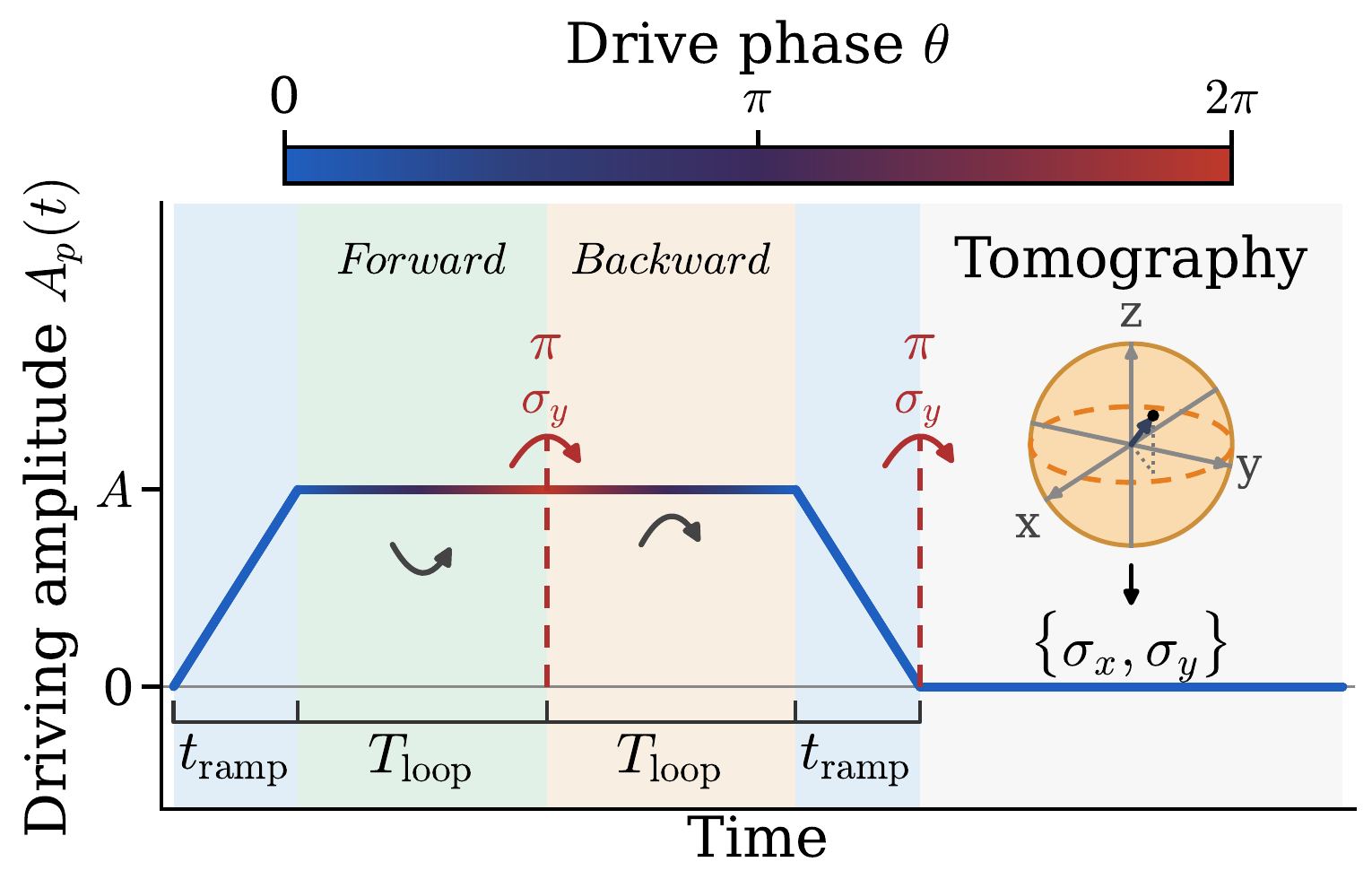}}
    \caption{Protocol for the cancellation of the dynamical phase. After preparation of the
    qubit in the ground state, the amplitude is ramped up with driving phase $\theta=0$. After time $t_\text{ramp}$, the amplitude is kept constant, while the drive phase $\theta$ is adiabatically increased during $N$ driving periods until it reaches the value $2\pi$. Next, a $\sigma_y$-pulse is applied, and the parameters are tuned back in the same way. A final $\sigma_y$-pulse restores the ground state. For Ramsey interference, a $\sigma_y$-pulse is applied initially to create a superposition between ground and excited state. Finally, the populations are measured by tomography.} 
\label{fig:protocol}
\end{figure}

Combining the preparation and the spin-echo idea leads to the protocol
sketched in Fig.~\ref{fig:protocol}.  It starts with a preparation in the
ground state of the undriven system for a given value of the detuning
$\epsilon$. Then during a time $t_\text{ramp}$ the amplitude $A_p(t)$ is increased from zero
to $A$ so that the wavefunction becomes proportional to the
Floquet mode $|\phi_-(0)\rangle$ for $\epsilon$ and $A$.  Next, the phase
$\theta$ of the driving is adiabatically tuned from $0$ to $2\pi$.
According to Eqs.~\eqref{eq:phidot} and \eqref{eq:equivalence}, the
wavefunction acquires a dynamical phase plus the AA phase.  After
applying a $\sigma_y$-pulse, the procedure is inverted, which
compensates the dynamical phase $-qNT$ and provides the AA phase a second time.  A
final $\sigma_y$-pulse brings the system back to its ground state.
In total, $|\psi_\text{final}\rangle = e^{2i\gammaAA}
|\psi_\text{initial}\rangle$.

\sect{Non-adiabatic corrections}
Our analytic predictions are based on ideal adiabatic following to the Floquet states, which may require rather long protocol times.  Any experimental realization, however, is limited by its coherence time, which may be so short that non-adiabatic corrections play a role.
In general, adiabatic following a Floquet mode requires $\omega=\Omega/N_\text{loop}$ to be small compared to $\Delta q^2/ |\llangle \phi_+|\partial_\theta \mathcal{H}|\phi_-\rrangle|$, where the outer angular brackets denote time average and $\Delta q$ the local quasienergy gap \cite{HonePRA97,WeinbergPR17}. In the present case, we benefit from the fact that $\mathcal{H}$ depends on $t$ and $\theta$ through $t+\theta/\Omega$, so that the relevant coupling $\partial_\theta \mathcal{H}$ is proportional to $\Delta q$. Then the condition reduces to $\omega\lesssim\Delta q$, so that the required $N_{\text{loop}}$ grows as $\Delta q^{-1}$. The remaining matrix element is proportional to the driving amplitude, so that the largest amplitude sets the limit.  For a derivation, see the \cite{supplement}.

\begin{figure}
\centerline{\includegraphics[width=\columnwidth]{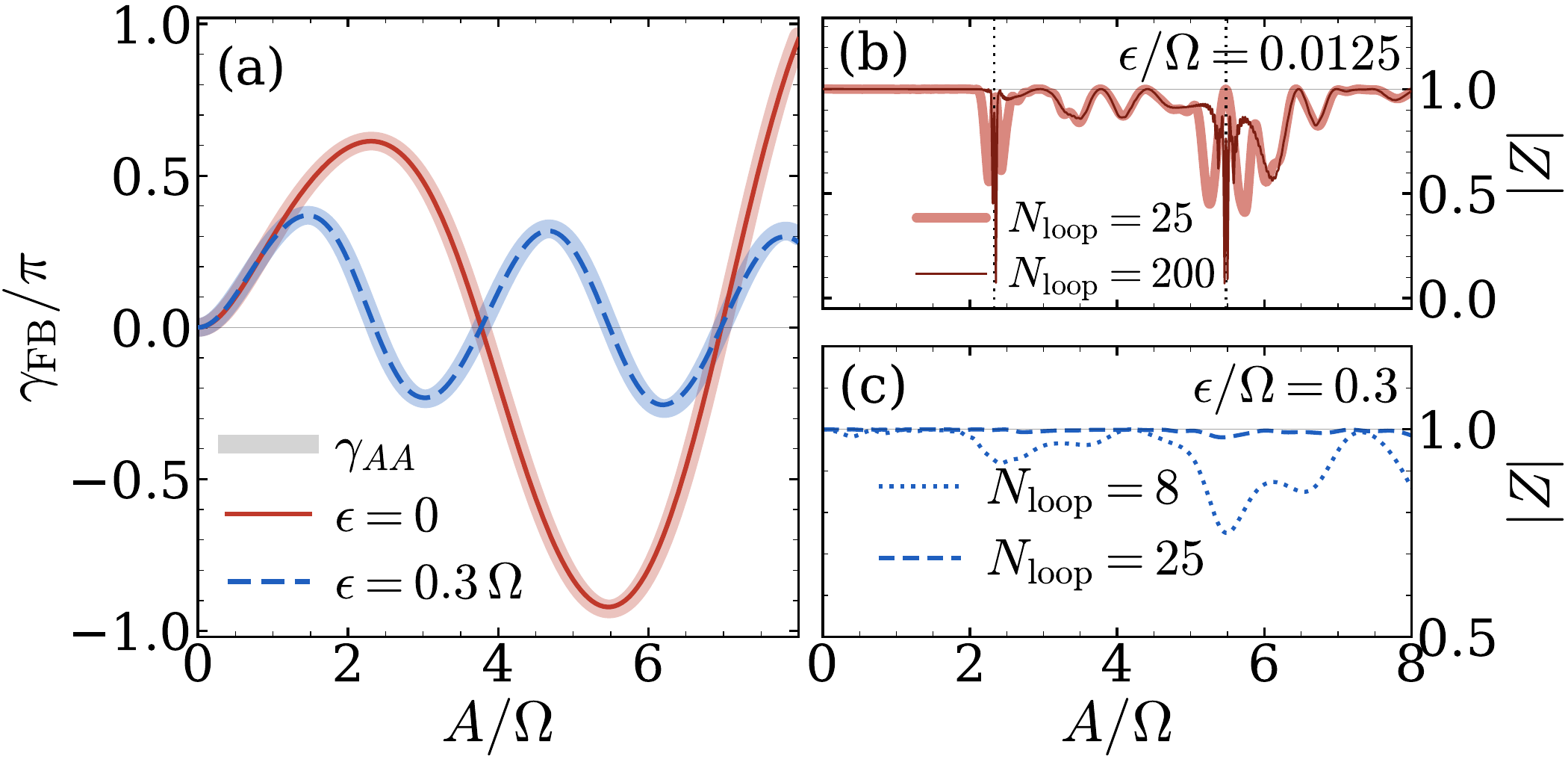}}
    \caption{Simulation of the protocol sketched in Fig.~\ref{fig:protocol} with preparation in the ground state for tunneling $\Delta=\Omega/2$ and various detunings.
    (a) Phase of the overlap $Z$ (thin lines) compared to the theoretically expected value $2\gammaFB$ [broad matte line of same color, cf.\ Fig.~\ref{fig:quasi}(a)] for protocol time $T_\text{loop}=25T$. The detunings are $\epsilon=0$ (red) and $0.3\Omega$ (blue).
    (b,c) Loop fidelity $|Z|$ for detunings $\epsilon=0.0125\Omega$ (b) and $0.3\Omega$ (c) for various protocol times $T_\text{loop}=N_\text{loop}T$. The vertical dotted lines mark positions of the quasienergy crossings in the absence of detuning.} 
\label{fig:ground_results}
\end{figure}

For an estimate of the relevance of non-adiabatic corrections, we simulate for the TLS in Eq.~\eqref{Hamiltonian} the protocol with various $T_\text{loop} = N_\text{loop}T$.  The computation starts with the preparation in the ground state for a given value of the detuning, $|\psi_\text{initial}\rangle = |g\rangle$.  Our central quantity of interest is the overlap $Z = \langle\psi_\text{initial}|\psi_\text{final}\rangle$, which for perfect adiabatic following has unit modulus while its phase is $2\gammaAA$.
Figure~\ref{fig:ground_results}(a) shows that already for the relatively short protocol time $T_\text{loop} = 25\,T$, its phase is practically indistinguishable from the ideal value $2\gammaAA$.  This is even the case for $\epsilon=0$ and amplitudes $A\gtrsim 2.4\Omega$, for which $A_p(t)$ passes through exact quasienergy crossings, see Fig.~\ref{fig:quasi}(b).

This seeming contradiction can be explained by a spatio-temporal symmetry for $\epsilon=0$.  Then the Hamiltonian \eqref{Hamiltonian} is invariant under the generalized parity $\mathcal{P}=\sigma_x\otimes(t\to t+T/2)$ \cite{PeresPRL91}, such that the Floquet modes can be classified as even and odd. Quasienergies of states with opposite parities may form exact crossings.  During ramp-up, non-adiabatic transitions can be attributed to $\dot A_p(t)\sigma_z\cos(\Omega t)$, which is also invariant under generalized parity. Consequently, the matrix element between the Floquet modes of the exact crossing vanishes, so that transitions can be caused only between replicas of the Floquet modes. Then the relevant minimal gap is of order $\Omega$ and non-adiabatic corrections are minor.

In an experiment, it will be difficult to keep the system at exactly zero detuning, such that the benefits of the generalized parity may be limited.
Furthermore, as we will see below, a finite detuning will allow to increase fidelity. For that case, the minimal splitting becomes $\Delta q\simeq \epsilon$, while adiabatic following requires $N_\text{loop}\gtrsim\Omega/\epsilon$ \cite{supplement}.
In Fig.~\ref{fig:ground_results}(b), we test adiabaticity for such a small detuning $\epsilon=\Omega/80$.  As a criterion, we use the loop fidelity $|Z|$, which turned out to be more sensitive to corrections than the phase.  This reveals that for $T_\text{loop}=25T$, non-adiabatic transitions play a significant role.
Even for the significantly longer duration $T_\text{loop}=200T$, the fidelity may be as low as $|Z|\sim 0.1$.
The solution is to maintain some distance to the crossing by increasing the detuning. Figure~\ref{fig:ground_results}(c) shows that for $\epsilon=0.3\Omega$, good adiabatic following can already be achieved with $N_\text{loop}=25$.

\sect{Ramsey interference}
With the protocol developed so far, the wavefunction acquires a global phase, which is not experimentally accessible, while any measurement requires a phase reference.  We have already seen in Fig.~\ref{fig:quasi} that for our TLS, the non-equivalent Floquet modes have opposite phases (a formal proof is based on the chirality of the Floquet Hamiltonian $H(t)-i\partial_t$ \cite{supplement}). Therefore, we can extend the basic protocol with Ramsey interferometry \cite{RamseyRMP90}. The idea is to create a superposition of the two states by a suitable $\pi/4$-pulse, such that the relative phase determines the final populations.

Again we start with the ground state $|g\rangle$ of the undriven Hamiltonian $H_0 = \Delta\sigma_x/2 + \epsilon\sigma_z/2$.  The Ramsey pulse $R = e^{i\pi\sigma_y/4}$ creates a superposition $\propto |g\rangle+|e\rangle$, which by the above protocol evolves into $e^{2i\gammaAA}|g\rangle + e^{-2i\gammaAA}|e\rangle$.  The outcome of a further Ramsey pulse is $|\psi_\text{final}\rangle = i\sin(2\gammaAA)|g\rangle + \cos(2\gammaAA)|e\rangle$.  A final tomography will find the TLS in the excited state with probability
\begin{equation}
    P_e = \cos^2(2\gammaAA) = \tfrac{1}{2}[1 + \cos(4\gammaAA)] .
    \label{Pe}
\end{equation}

\begin{figure}
\centerline{\includegraphics[width=\columnwidth]{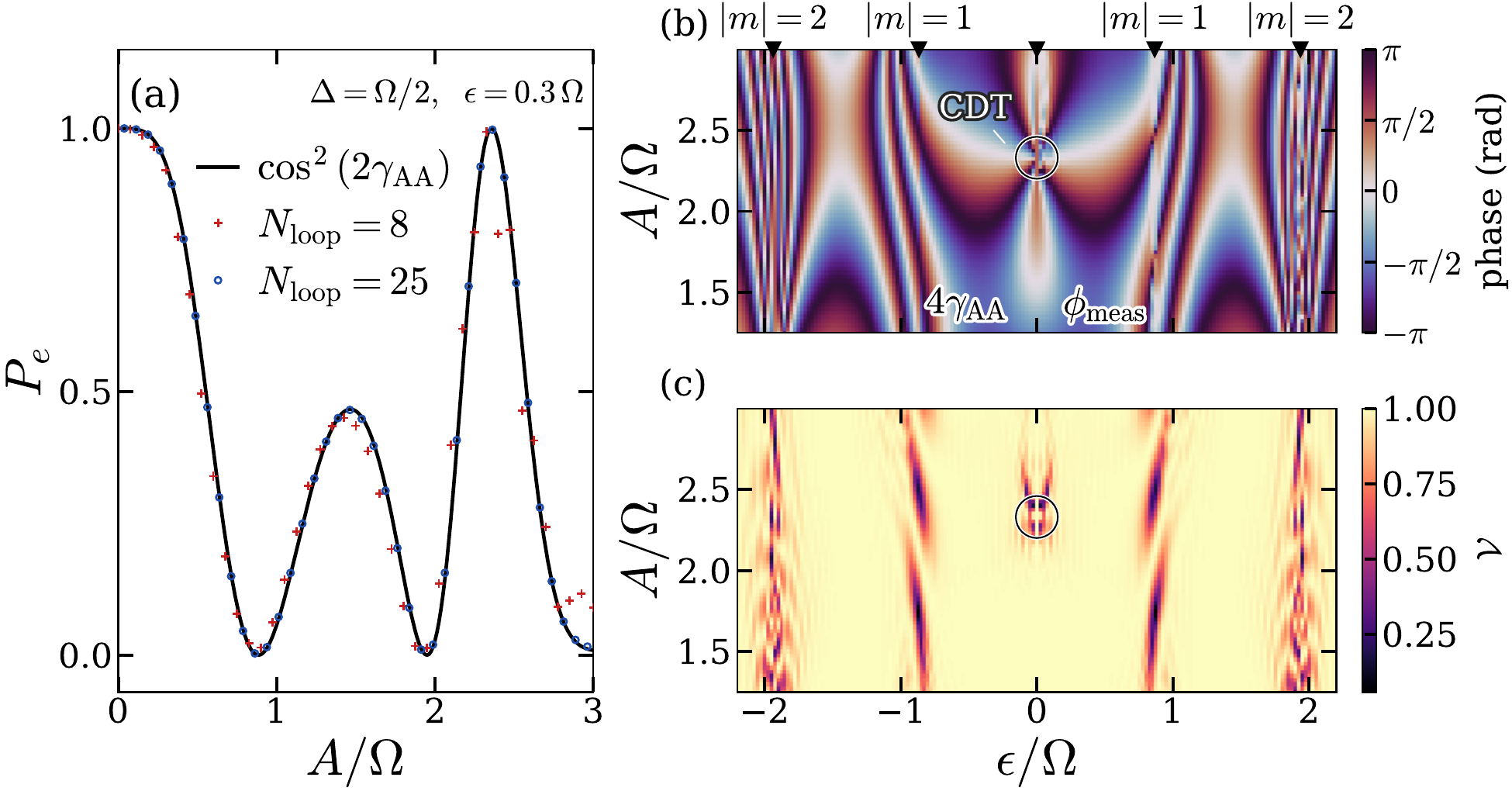}}
    \caption{%
    (a) Population of the excited state after the protocol with Ramsey pulses for various loop times (symbols) compared to the expectation in Eq.~\eqref{Pe}. Tunneling and detuning are $\Delta=0.5\Omega$ and $\epsilon=0.3\Omega$, respectively.
    (b) Theoretically expected value $\gammaAA$ from Eq.~\eqref{anandan_formula} (left half) compared to the simulated measurement (right half). The result is independent of the sign of $\epsilon$ due to the symmetry relation $\sigma_x H(A,\epsilon,\theta)\sigma_x=H(A,-\epsilon,\theta+\pi)$.
    (c) Visibility of the Ramsey pattern for the simulated measured phase as a function of the detuning and amplitude for $N_\text{loop}=25$.}
\label{fig:experimental}
\end{figure}

To demonstrate the proper operation for finite protocol times $T_\text{loop}$, we repeat our simulation with Ramsey pulses included and with a focus on observables.  We have already seen that a significant detuning improves adiabatic following, so we restrict ourselves to $\epsilon=0.3\Omega$.  Figure~\ref{fig:experimental}(a) shows the excitation probability from the simulation, $P_e = |\langle e|\psi_\text{final}\rangle|^2$, for various loop times compared to the conjecture in Eq.~\eqref{Pe}. It shows that already $N_\text{loop}= 25$ is sufficient for a good agreement. An important point is the visibility of the fringes.  For the present detuning, the AA phase varies only on the order of $\pi/2$ [see Fig.~\ref{fig:quasi}(a)], while the factor 4 in Eq.~\eqref{Pe} enhances the effect such that $P_e$ gives rise to full visibility from zero to unity. 

Figure~\ref{fig:experimental}(b) shows the difference between the simulated protocol phase and its predicted counterpart from Eq.~\eqref{anandan_formula} over the ($A,\epsilon$) plane. Only when the detuning is tiny or when it comes close to the one-photon resonance with $E_e-E_g=\Omega$, do we notice a relevant difference. In a possible experiment, these regions may be avoided by an adiabatic ramp-up with a time-dependent detuning.
Finally, Fig.~\ref{fig:experimental}(c) shows the visibility of the protocol phase. A relevant visibility loss is found around the same points as in panel (b). 
Nevertheless, it can still approach values close to unity at points where the simulated and the predicted phases do not match.

Our protocol is similar to the one used in the experiment of Ref.~\cite{LeekS07} for the measurement of an (adiabatic) Berry phase. It may be implemented with the same setup. The experiment considered resonant driving with a small amplitude, such that a rotating-wave approximation allows a mapping to an effective Hamiltonian with merely adiabatic time dependence.  For our generalization to the Floquet regime with arbitrary driving frequency and amplitude, one instead measures the non-adiabatic AA phase \cite{AharonovPRL87}.  This indicates that within the rotating-wave approximation, the AA phase and the FB phase must be identical to the adiabatic Berry phase of the rotating-wave Hamiltonian.  This is indeed the case and identifies Ref.~\cite{LeekS07} as a limiting case of our protocol, as we demonstrate in the \cite{supplement}.

\sect{Conclusions}
We have developed a protocol for the measurement of the geometric AA phases of the Floquet states of a strongly driven quantum system. It is based on the equivalence of this phase and a FB phase that emerges by slowly changing the phase of the driving field from $0$ to $2\pi$. Remarkably, the non-adiabatic AA phase can be measured by adiabatic parameter variation. An essential ingredient is the chirality of the Hamiltonian, which is required for the cancellation of a dynamical phase.

The protocol can be considered as a non-adiabatic generalization of the experiment by Leek \textit{et al.}\ \cite{LeekS07} in which an (adiabatic) Berry phase of a rotating-wave Hamiltonian has been measured.  That protocol emerged as a limiting case of ours.  The requirements for an implementation are practically the same. In particular, no sophisticated Floquet state preparation is needed, as it is sufficient to start in the ground state of the Hamiltonian in the absence of the driving.  A possible dynamical phase acquired during the ramp-up of the driving amplitude is canceled in the same way as the one emerging during the central part of the protocol.

To demonstrate that non-adiabatic corrections are not an issue, we have simulated the complete protocol from the preparation to a final tomography.  It turned out that even for the operation within less than 50 driving periods, non-adiabatic transitions stay at a tolerable level. This is particularly relevant to avoid coherence loss and to reduce the impact of insufficient parameter control.

The scheme should also work for systems beyond a TLS. While implementing the variation of the drive phase may be achieved straightforwardly for any system, feasibility will depend on a sufficiently stable spatio-temporal chirality. Further investigation may explore the limits set by environmental noise and imperfect parameter control.

\begin{acknowledgments}
This work was supported by the Spanish Ministry of Science, Innovation, and Universities under Grant Nos.\ PID2023-149072NB-I00 and AIA2025-163435-C44, and through a FPI grant (CMF).
\end{acknowledgments}

\appendix
\section{Aharonov-Anandan phase of the Floquet states}
\subsection{General case}
We start from the general expression of the Anandan phase of a $T$-periodic state \cite{AharonovPRL87},
\begin{equation}
    \gammaAA = \int_0^T dt \langle \phi(t)| i\partial_t |\phi(t)\rangle .
    \label{AAphase}
\end{equation} 
The Fourier representation $|\phi(t)\rangle=\sum_n e^{-in\Omega t}|\phi_{n}\rangle$ with the normalization $\sum_n\langle\phi_{n}|\phi_{n}\rangle=1$ leads to
\begin{align}
    \gammaAA
    ={}&\int_0^T dt \sum_{n,n'} e^{in'\Omega t}\langle\phi_{n'}| i \partial_t  e^{-in\Omega t}|\phi_{n}\rangle \\
    ={}& 2\pi \sum_n  n\langle\phi_{n}|\phi_{n}\rangle.
\end{align}

For the particular case of a Floquet state, the corresponding solution of the Schr\"odinger equation reads $|\psi(t)\rangle = e^{-iqt}|\phi(t)\rangle$ with the quasienergy $q$.  Its total phase acquired during a driving period is $-qT$.  Substituting in Eq.~\eqref{AAphase} the time derivative via the Floquet Hamiltonian $\mathcal{H}=H(t)-i\partial_t$ and using the fact that $q$ is its eigenvalue yields $\gammaAA=(\bar{E}-q )T$, with the time-averaged energy expectation value, the mean energy $\bar E$ \cite{MooreJPA90b}.

By introducing the scaled time $\theta=\Omega t$, the Floquet operator can be written in the form $H(\theta)-i\Omega\partial_\theta$, which depends on the driving frequency only through the second term.  This allows one to apply the Hellmann-Feynman theorem which relates the derivative of an eigenvalue and the expectation value of the derivative of the corresponding operator. The application of the theorem is based on the interpretation of the Floquet operator as a genuine Hermitian eigenvalue problem in Sambe space, so that quasienergies behave like static eigenvalues. 

Therefore, the $\Omega$-derivative of the quasienergy yields
\begin{equation}
    \frac{\partial q}{\partial \Omega}= -\langle \langle \phi | i\partial_\theta |\phi\rangle \rangle,
\end{equation}
for fixed $A,\Delta,\epsilon$.
Together with Eq.~\eqref{AAphase} follows that the derivative of the quasienergy obeys~\cite{FainshteinJPB78},
\begin{equation}
    \frac{\partial q}{\partial \Omega} = -\gammaAA/2\pi.
    \label{eq:hellmann}
\end{equation}
This relates the geometric phase of a driven qubit to the change of the quasienergy with respect to the driving frequency.

\subsection{High frequency limit}
Analytical insight is a useful tool in order to describe phenomena and have a better understanding of the ongoing physics. Here we derive analytical expressions for the AA phase in the high-frequency regime $\Delta\ll\Omega$ for the two-level Hamiltonian
\begin{equation}
H(t)=\tfrac12\Delta\,\sigma_x+\tfrac12 \left[A\cos(\Omega t)+\epsilon\right] \sigma_z.
\label{eq:H}
\end{equation}
We start with a transformation with respect to the driving,
\begin{equation}
U(t)=\exp\left({-i\alpha(t)\sigma_z/2}\right),
\quad \alpha(t)=(A/\Omega)\sin(\Omega t),
\label{eq:U}
\end{equation}
by which we find the interaction-picture Hamiltonian,
\begin{align}
\tilde H(t)
={}& U^{\dagger}HU-iU^{\dagger}\dot U
\\
={}&\frac{\epsilon}{2}\sigma_z
+\frac{\Delta}{2}\left[\cos\alpha(t)\,\sigma_x-\sin\alpha(t)\,\sigma_y\right].
\label{eq:Htilde}
\end{align}
For a sufficiently large driving frequency, the dynamics is much slower than the oscillation of matrix elements of $\tilde H$. Then, within time-scale separation, we replace the Hamiltonian by its time average to obtain the time-independent effective Hamiltonian
\begin{equation}
H_\text{eff} = \frac{\epsilon}{2}\sigma_z+\frac{\Delta}{2} J_0(A/\Omega)\sigma_x 
\label{eq:Hm}
\end{equation}
with the eigenvalues $\pm\Delta_\text{eff}/2$, where
\begin{equation}
\Delta_{\text{eff}}=\sqrt{\epsilon^{2}+\Delta^{2}J_0^{2}(A/\Omega)},
\label{eq:quasienergies}
\end{equation}
which is correct up to order $\mathcal{O}(\Delta^3/\Omega^3)$ and also requires $\epsilon$ to be sufficiently small.

The back transformation of the eigenstates $|u_\pm\rangle$ provides the Floquet states $U(t)|u_\pm\rangle$ and reveals that the quasienergies are $q_\pm = \pm\Delta_\text{eff}/2$.

The corresponding AA phase follows readily from Eq.~\eqref{eq:hellmann} and reads
\begin{equation}
\gammaAA^{\pm}= \mp \frac{\pi\Delta A}{\Omega^2} J_1\left(A/\Omega \right) \frac{\Delta J_0(A/\Omega)}{\sqrt{\epsilon^2+\Delta^2J_0^2(A/\Omega)}} ,
    \label{eq:hf:det}
\end{equation}
where we have used the relation for the derivative of the Bessel function, $J_0'(x)=-J_1(x)$.
In the absence of detuning, $\epsilon=0$, the last factor becomes $\pm1$ such that
\begin{equation}
\gammaAA^{\pm}
=\mp\frac{\pi\Delta A}{\Omega^{2}}\,J_1(A/\Omega).
\label{eq:hf-result}
\end{equation}
Both expressions vanish when $A/\Omega$ equals a root of the Bessel function $J_1$.  The expression for $\epsilon\neq0$, in addition, vanishes at roots of $J_0$, even for arbitrarily small detuning.  

This qualitative difference explains an observation made in the context of Fig.~\ref{fig:quasi}(a) of the main text.
For non-zero detuning $\epsilon$, $\gammaAA$ acquires an additional zero between consecutive zeros of the $\epsilon=0$ case. 
The drive generates the exact micromotion $U(t)=e^{-i\frac{A}{2\Omega}\sin(\Omega t)\sigma_z}$, a rotation about $\hat{z}$, while the effective Hamiltonian $H_0=\frac{\epsilon}{2}\sigma_z+\frac{\Delta}{2}J_0(A/\Omega)\sigma_x$ governs the stroboscopic dynamics.

The common zeros are given by $J_1(A/\Omega)=0$, where the geometric phase vanishes independently of $\epsilon$. At $J_0(A/\Omega)=0$, with $\epsilon\neq0$, $H_0$ aligns the Floquet state with the $\hat{z}$-axis. The Bloch vector then becomes stationary and no solid angle is enclosed, so that $\gammaAA=0$ and $\bar{E}=q$. For $\epsilon=0$, instead, $H_0\propto\sigma_x$ and the state remains on the $\hat{x}$-axis of the Bloch sphere, maximally transverse to the rotation axis, so $|\gammaAA|$ is maximal. As $J_1=-J_0'$, within the high-frequency approximation, the zeros of the AA phases are independent of $\epsilon$, as observed in the main text.

\section{Equivalence of $\gammaFB$ and $\gammaAA$}

For a Hamiltonian that depends on a parameter $\theta$, the Floquet equation reads
\begin{equation}
    \big(  H(t,\theta)-i\partial_t\big) |\phi(t,\theta)\rangle=q(\theta)|\phi(t,\theta)\rangle.
    \label{FlEq}
\end{equation}
Its quasienergies do not depend on time or an initial phase. If, like in the present case, the parameter is the phase of the driving field, the parameter dependence of the Hamiltonian corresponds to a time shift $t\to t+\theta/\Omega$.  Therefore, it is invariant under a transformation $(t,\theta)\rightarrow (t+\tau,\theta-\Omega\tau)$. Hence, both $|\phi(t,\theta)\rangle$ and $|\phi(t+\tau,\theta-\Omega\tau)\rangle$ are solutions of the Schrödinger equation with the same quasienergy.

This implies that both states are related by a phase transformation and as the overall phase of a Floquet state is a gauge degree of freedom  we are free to choose a gauge such that
$|\phi(t,\theta)\rangle=|\phi(0,\theta+\Omega t)\rangle$. This gauge is single-valued provided $|\phi(0,\theta)\rangle$ is $2\pi$-periodic in $\theta$. Consequently, $\partial_\theta|\phi(t,\theta)\rangle=\Omega^{-1}\partial_t|\phi(t,\theta)\rangle$. Therefore, we find an equivalence of time evolution and phase shifts for Floquet states. This establishes the key relation for studying the AA phase of a Floquet state by varying its driving phase.

For slow parameter variation, the dynamics is given by adiabatic following of Floquet states \cite{WeinbergPR17}, such that the remaining freedom is the phase in the ansatz,
\begin{equation}
    |\psi(t)\rangle=e^{i\varphi(t)}|\phi(t,\theta(t))\rangle.
\end{equation}
Inserting into the Schrödinger equation, we obtain
\begin{equation}
i\partial_t|\psi(t)\rangle=e^{i\varphi(t)}\left[ -\dot{\varphi}|\phi\rangle+i\dot{\theta}\partial_\theta|\phi\rangle+i\partial_t|\phi\rangle   \right]    .
\end{equation}
The time derivative can be substituted by Eq.~\eqref{FlEq}, such that projecting onto 
$\langle\psi(t)|$ and neglecting non-adiabatic transitions yields
\begin{equation}
    \dot{\varphi}=-q+\dot{\theta}\langle \phi|i\partial_\theta|\phi\rangle.
    \label{eq:ansatz}
\end{equation}
This resembles the expression for adiabatic following without ac driving, but now with the adiabatic energy replaced by the quasienergy. Let us integrate the different terms in Eq.~\eqref{eq:ansatz} for $t=0\ldots NT$. 
The equivalence of time and phase shifts ensures that the quasienergy is time independent, so integrating this term simply yields $-q NT$. 
For the geometric contribution, we must analyze the Floquet state's explicit and implicit time dependences as they belong to different time scales:
slowly through $\theta(t)$ fast by the explicit $T$-per intervals $I_n=\left[ (n-1)T,nT \right]$ with $n=1,\dots,N$. Within one period, $\theta$ advances only by
\begin{equation}
    \delta \theta=\dot{\theta}T= 2\pi/N \ll 1,
\end{equation}
hence within time-scale separation, we can assume that it freezes to a constant value $\theta_n = \theta(nT)$. Since $\langle \phi|i\partial_\theta|\phi\rangle$ is $T$-periodic in $t$, all integrals are the same, and a discretization is admitted. The Floquet-Berry phase acquired during the cycle is
\begin{equation}
    \gammaFB=\sum_{n=1}^N \frac{\theta_n-\theta_{n-1}}{T} \int_0^T dt \langle \phi(t,\theta_n) |i\partial_\theta|\phi(t,\theta_n)\rangle,
\end{equation}
where the time derivative $\dot{\theta}$ has been discretized as $\dot\theta = (\theta_n-\theta_{n-1})/T$.  The remaining integral provides the period-averaged connection
\begin{equation}
    \mathcal{A}(\theta)\equiv \frac{1}{T} \int_0^T dt~\langle\phi(t,\theta)|i\partial_\theta|\phi(t,\theta)\rangle
    =\gammaAA/2\pi,
\end{equation}
such that
\begin{equation}
\gammaFB = \gammaAA [\theta(NT)-\theta(0)] / 2\pi.
\end{equation}

\section{Chirality and cancellation of the dynamical phase}

The Hamiltonian in Eq.~\eqref{Hamiltonian} of the main text obeys $\sigma_y H(t,\theta)\sigma_y=-H(-t,-\theta)$. The anti-commutation relation of the Pauli matrices and the symmetry of the cosine provide the chirality of the Floquet operator, $S=\sigma_y \otimes(t,\theta\rightarrow -t,-\theta)$, which inverts the sign of the time derivative such that
\begin{equation}
    S\mathcal{H}(\theta)S^{-1}=-\mathcal{H}(-\theta).
    \label{eq:chirality}
\end{equation}
The chirality links Floquet states with opposite quasienergy as $|\phi_\mp(t,\theta)\rangle\propto\sigma_y|\phi_\pm(-t,-\theta)\rangle$, where $-q_-=q_+$. Since $S^2=\mathbb{1}$, one can fix the gauge such that
\begin{equation}
    |\phi_+(t,\theta)\rangle=\sigma_y|\phi_-(-t,-\theta)\rangle .
    \label{eq:chir_gauge}
\end{equation}
Inserting Eq.~\eqref{eq:chir_gauge} into Eq.~\eqref{AAphase} and substituting $t\to -t$ gives $\gammaAA^{(+)}=-\gammaAA^{(-)}$, consistent with the quasienergy swap. By $T$-periodicity, $|\phi(-kT)\rangle=|\phi(kT)\rangle$ for integer $k$ and any value of $\theta$. Therefore, a $\sigma_y$ pulse maps $|\phi_-\rangle\rightarrow|\phi_+\rangle$ up to a global phase factor.
For a loop starting in $|\phi_-\rangle$, the echo sequence gives:
\begin{align}
    \textbf{(i)}\;(\phi_-,~\theta:0\rightarrow2\pi)
    :\quad & -q_-NT+\gammaAA^{(-)},\\[2mm]
    \textbf{(ii)}\;(\phi_+,~\theta:2\pi\rightarrow0)
    :\quad & -q_+NT-\gammaAA^{(+)} \nonumber\\
    &= +q_-NT+\gammaAA^{(-)}.
\end{align}
Then the dynamical phases cancel each other, while the geometric phases add up to $2\gammaAA^{(-)}$. The reversed phase change of the second loop is necessary because the chirality $S$ implies $\theta\rightarrow-\theta$, see Eq.~\eqref{eq:chirality}. The dynamical phase of the ramp-up on $|\phi_-\rangle$ and the time-mirrored ramp-down on $|\phi_+\rangle$ cancel for the same reason.  In the Ramsey protocol, both Floquet states are transported simultaneously and acquire a relative phase $\pm2\gammaAA$ per loop, hence, $\phi_{\text{meas}}=4\gammaAA$.

\section{Readout and Ramsey pulses}
In the presence of a fixed detuning, the static part of the Hamiltonian~\eqref{Hamiltonian} becomes
\begin{equation}
    H_0=\frac{1}{2}\left(  \Delta \sigma_x+\epsilon\sigma_z   \right)=\frac{|\hat{n}|}{2} \hat{n}\cdot \vec{\sigma},
\end{equation}
With $\hat{n}=(\cos\alpha,0,\sin\alpha)$ and $\alpha=\arctan(\epsilon/\Delta)$, so that the eigenstates are tilted away from the $x$ axis by an angle $\alpha$. The rotation $R_\alpha=e^{i\alpha\sigma_y/2}$ satisfies
\begin{equation}
R_\alpha\sigma_xR_\alpha^\dagger=\hat{n}\cdot\vec{\sigma},
\end{equation}
so the eigenstates of $H_0$ are $R_\alpha|\pm_x\rangle$, and the adiabatic ramp turns $R_\alpha|-_x\rangle$ into the corresponding Floquet mode. 

The key point is that $\hat{n}$ lies in the $xz$-plane, i.e., perpendicular to the rotation axis of the Ramsey pulse $R=e^{i\pi\sigma_y/4}$. Then a $\pi/2$-rotation about the $\hat{y}$-axis maps $\hat{n}$ onto a direction orthogonal to $\hat{n}$ it for any $\alpha$, such that $R$ produces an equal superposition of the eigenstates of $H_0$ and the tilt need not be considered for the state preparation. Equivalently, $[R,R_\alpha]=0$. The role of the Ramsey pulse is interferometric, as it converts a relative phase into populations.
Moreover, since $\sigma_y$ anticommutes with $\sigma_{x,z}$, the chirality relation $\{\sigma_y,H(t)\}=0$ as well as the cancellation of the dynamical phase are preserved.

Only the readout is affected by the tilt. The transverse components must be measured relative to $\hat{n}$, i.e., along $\sigma_y$ and $R_\alpha\sigma_zR_\alpha^\dagger$, or equivalently via the populations after the second Ramsey pulse, Eq.~\eqref{Pe} of the main text. Since both rotations are generated by $\sigma_y$, the basis alignment and the closing Ramsey pulse merge into a single pulse $R_\alpha R^\dagger=e^{-i(\pi/2-\alpha)\sigma_y/2}$. A detuning only shifts the angle of the last pulse by some known amount $\alpha$.

Lastly, let us comment on a detail of the second Ramsey pulse. While theoretically one would like to apply $R^\dagger$, the natural experimental choice is to apply $R$ again. The composed transformation $R^2=e^{i\pi\sigma_y/2}=i\sigma_y$ is a full $\pi$-rotation, i.e., a bit flip, while $RR^\dagger=1$. The difference is a mere shift of the Ramsey pattern, i.e., of the calibration point according to
\begin{equation}
    \gammaAA=0:~P_e^{(R,R)}=1,~~P_e^{(R,R^\dagger)}=0 ,
\end{equation}
while the non-trivial part of the fringe remains $\cos(4\gammaAA)$. In the main text, we choose the former, so that
\begin{equation}
    P_e=\cos^2(2\gammaAA)=\frac{1}{2}[1+\cos(4\gammaAA)].
\end{equation}

\section{Condition for adiabatic following}

\begin{figure}
\centerline{\includegraphics[width=0.7\columnwidth]{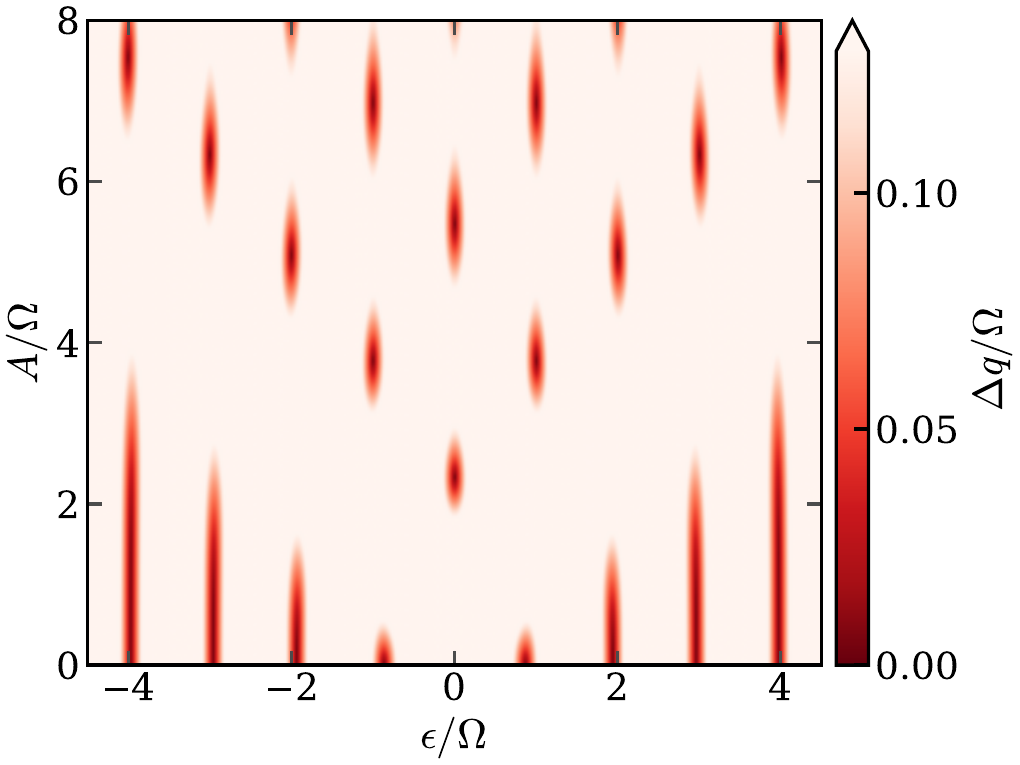}}
    \caption{Quasienergy gap $\Delta q$ in the ($\epsilon,A$) plane. The dark areas mark accidental degeneracies found for integer detuning \cite{KohlerQ26} and at $m$-photon resonances for $A=0$. There, adiabatic following breaks down.}
\label{fig:sup:quasimap}
\end{figure}

Adiabatic following is crucial for our protocol. Here we elaborate on the conditions under which this can be expected. The fidelity of the protocol is controlled by Floquet adiabatic perturbation theory \cite{WeinbergPR17,HonePRA97}, the generalization of adiabatic perturbation theory to periodically driven systems. Under the assumptions of a well-defined Floquet Hamiltonian and a non-degenerate target Floquet mode; both satisfied in a finite-dimensional system away from exact crossings. For a slow parameter $\lambda(t)$, the adiabaticity condition reads~\cite{WeinbergPR17}:
\begin{equation}
    \eta_{\alpha\beta}^{(\ell)}=\frac{|\dot{\lambda} \langle\langle \phi_\beta^{(\ell)}| \partial_\lambda \mathcal{H}|\phi_\alpha\rangle \rangle|  }{(q_\beta-q_\alpha+\ell\Omega)^2}  \ll1, 
  \qquad\forall\,\ell\in\mathbb{Z},\;\beta\neq\alpha  .
  \label{eq:ad}
\end{equation}
The smallest denominator sets the limit, which for a two-level system is the minimal quasienergy gap
\begin{equation}
  \Delta q = \min_{\ell\in\mathbb{Z}}\bigl|q_+-q_-+\ell\Omega\bigr| \in [0,\Omega/2].
  \label{eq:folded}
\end{equation}
Arbitrarily close to a quasienergy crossing, non-adiabatic leakage is unavoidable. There, we have to consider two cases:
\begin{itemize}
    \item \textbf{Coherent destruction of tunneling (CDT) \cite{GrossmannPRL91}}. $\ell=0$, $\Delta q=q_+-q_-\rightarrow0$, so $\eta^{(0)}\rightarrow\infty$.   
    \item \textbf{Landau-Zener-Stückelberg (LZS) and multiphoton resonances \cite{ShevchenkoPR10}}.  $\ell=m\neq0$, where the unfolded splitting matches an integer number of drive quanta, $q_+-q_-\simeq m\Omega$, so the folded gap closes again.  
\end{itemize}
Figure~\ref{fig:sup:quasimap} shows the quasienergy gap in the ($\epsilon,A$) plane obtained from the Floquet spectrum alone. It locates the degeneracies over a wide parameter range, but does not quantify how much they degrade the measurement.  This has to be evaluated by a simulation of the full protocol.

For $\epsilon=0$, the system has parity protection, because the Hamiltonian is invariant under the generalized parity $\mathcal{P} = \sigma_x \otimes (t \rightarrow t + T/2)$ \cite{PeresPRL91}. Both $\partial_A \mathcal{H}$ and $\partial_\theta \mathcal{H}$ are $\mathcal{P}$-even. In Sambe space the replica $|\phi^{(\ell)}\rangle\rangle$ carries parity $(-1)^{\ell}p$, with $p=\pm1$ the parity of the associated state. Therefore, the matrix elements in Eq.~\eqref{eq:ad} between states of opposite total parity become zero and the coupling vanishes. 
At any finite detuning $\epsilon$, the coupling will not vanish, while the quasienergy gap remains small, so the problematic region is the immediate vicinity of the CDT points.

Multiphoton resonances occur at $\sqrt{\Delta^2+\epsilon^2}=m\Omega$, i.e., $\epsilon= \pm\sqrt{(m\Omega)^2-\Delta^2}$. Their position is independent of $A$, but their width scales as $|J_m(A/\Omega)|$. For $\epsilon=0$, $\Delta>\Omega$ would be needed for these resonances to appear, so it is not problematic for $\Delta=\Omega/2$.

Let us evaluate Eq.~\eqref{eq:ad} for the two slow processes of the protocol, namely ramp-up and phase shift, for which the adiabaticity condition has to be evaluated separately.  
\begin{itemize}
    \item \textbf{Phase shift.} For the drive phase $\theta=\omega_{\text{loop}} t$ with $\omega_{\text{loop}}=\Omega/N_{\text{loop}}$, the condition reads
    \begin{equation}
        \begin{gathered}
        \eta_\text{loop}=\frac{\omega_\text{loop} M_\theta}{(\Delta q)^2}=\frac{\Omega M_\theta} {N_\text{loop}\Delta q^2} \ll 1
       \\
        M_\theta= \langle\langle \phi_\beta ^{(\ell)}|\partial_\theta \mathcal{H} |\phi_\alpha\rangle\rangle .
        \end{gathered}
    \end{equation}
This can be simplified if we work in the extended Sambe space, so that $\partial_\theta \mathcal{H}=i[\mathcal{N},\mathcal{H}]$, where $\mathcal{N}$ is the photon number operator in the Floquet picture \cite{PerezGonzalez26}. Then, the matrix element becomes 
    \[
   M_\theta=\frac{i(q_\beta-q_\alpha+\ell\Omega)}{\Omega}\langle\langle \phi_\beta^{(\ell)}|H(t)|\phi_\alpha\rangle\rangle ,
    \]
    where $H(t)$ is the Hamiltonian, not the Floquet operator. One power of the gap cancels, and
    \begin{equation}
    N_\text{loop}\gg \frac{|\langle\langle \phi_\beta^{(\ell)}|H|\phi_\alpha\rangle\rangle |}{|q_\beta-q_\alpha+\ell\Omega|} .
    \end{equation}

    \item \textbf{Ramp.} For the linear ramp $A_p(t)=A t/t_\text{ramp}$, $\dot{A}_p=A/t_\text{ramp}$
    and $\partial_{A_p}\mathcal{H}=\tfrac{1}{2}\cos(\Omega t+\theta)\sigma_z$. No gap cancellation occurs and the condition for the ramp reads
\begin{gather}
  \frac{\Omega t_{\mathrm{ramp}}}{2\pi}
  \gg \max_{A_p\in[0,A]} \frac{\Omega\, A\, |M_A(A_p)|}{2\pi\, \Delta q(A_p)^2}
  \propto \Delta q_{\min}^{-2}, \nonumber \\
  M_A(A_p) = \langle\langle \phi_\beta^{(\ell)} | \partial_{A_p} \mathcal{H} | \phi_\alpha \rangle\rangle .
  \label{eq:Nramp}
\end{gather}
\end{itemize}
The ramp is limited by the smallest gap along $A_p\in[0,A]$. A first comparison of the adiabaticity limits lets us note that $N_{\text{loop}}$ scales as $\Delta q^{-1}$, while $N_{\text{ramp}}$ scales as $\Delta q^{-2}$. Despite the different coupling matrix elements, this scaling allows us to conclude that the condition for the ramp is less strict. Hence, our choice $N_\text{ramp}=N_\text{loop}$ is sufficient for adiabatic following also during ramp-up.

\begin{figure}
\centerline{\includegraphics[width=\columnwidth]{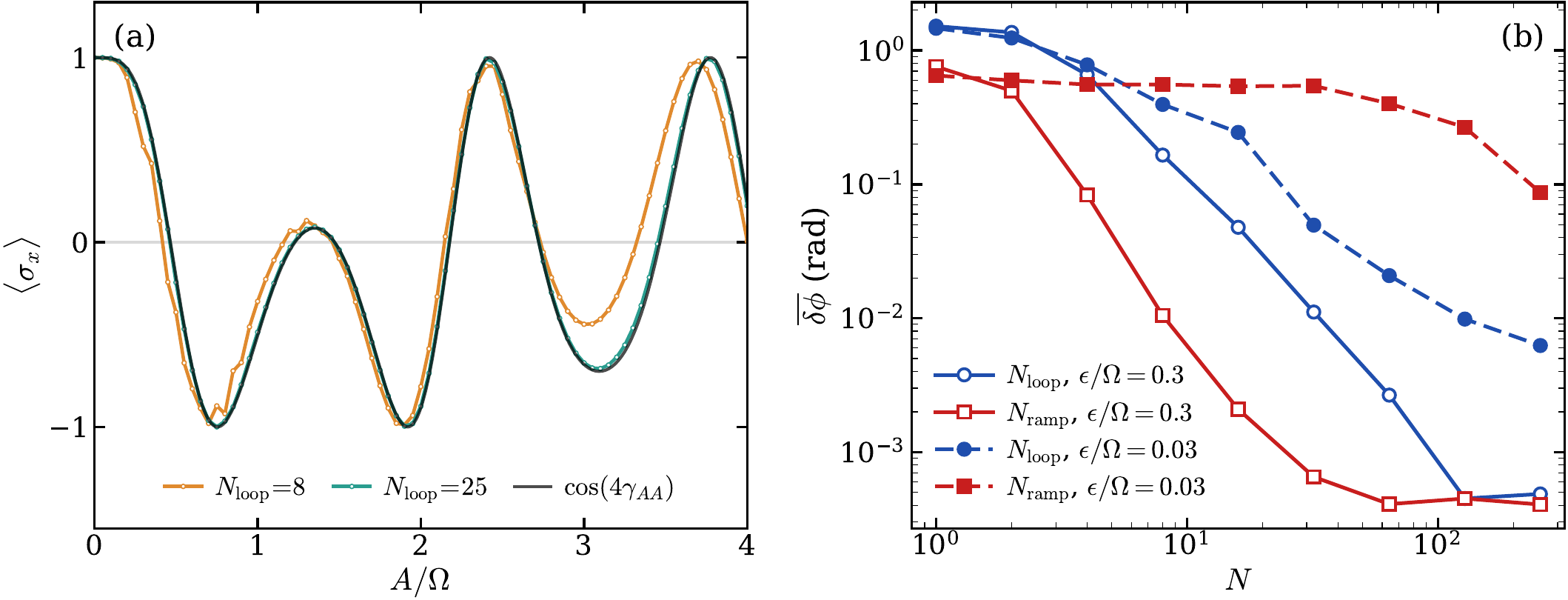}}
    \caption{(a) Protocol observable $\langle \sigma_x\rangle$ after the loops and the $\pi$-pulses as a function of the scaled amplitude $A/\Omega$ for the loop
durations $N_\text{loop}$ = 8 and 25, compared with the theoretical
value $\cos(4\gammaFB)$.
(b) Adiabaticity comparison between $N_\text{loop}$ (blue) and $N_\text{ramp}$ (red). While varying one parameter, the other is fixed at a value $N=256$ to avoid leakage from the other source. Results are shown for $\epsilon=0.3\Omega$ (solid lines) and $\epsilon=0.03\Omega$ (dashed).}
\label{fig:sup:ad}
\end{figure}

Figure~\ref{fig:sup:ad}(a) demonstrates the equivalence numerically: the observable $\langle \sigma_x\rangle=\cos(4\gammaFB)$ obtained from the simulated protocol is compared with the prediction of Eq.~\eqref{anandan_formula} from the Floquet modes. The convergence is obtained with growing $N_\text{loop}$; for $N_\text{loop} = 25$, the deviation becomes negligible over the full amplitude range. This demonstrates the equivalence between the FB phase from the adiabatic protocol and the non-adiabatic AA phase, while $N_\text{loop}=8$ exhibits the breakdown of the protocol when the driving phase is varied too fast. This confirms the fast convergence of the adiabaticity parameter used for the main text. 

Figure \ref{fig:sup:ad}(b) shows the phase error as a function of $N_\text{ramp}$ and $N_\text{loop}$. When varying one of these two parameters, we keep the other at the value $N=256$ to ensure adiabaticity within the respective part. 
The results for $\epsilon=0.3\Omega$ show that the condition on $N_\text{ramp}$ is less strict than the one for $N_\text{loop}$. In the main text, we use the same value for both, such that the ramp-up does not lead to any relevant non-adiabatic correction.

The only exception is when $A_p(t)$ comes close to a resonance, e.g., for small $\epsilon$ and an amplitude beyond the first quasienergy crossing, $A\gtrsim 2.4\Omega$.  Then $\Delta q^{2}$ appears in the denominator of the adiabaticity condition and may become critical.  This may be avoided by a modified protocol in which a variation of the detuning avoids coming close to the degeneracy point.

\section{Rotating-wave approximation}

Leek \textit{et al.}~\cite{LeekS07} started from a TLS Hamiltonian driven close to resonance with a relatively small amplitude, such that the model can be described within a rotating-wave approximation (RWA) by a time-independent Hamiltonian. They conjectured that the geometric phases acquired upon variation of the drive phase are the (adiabatic) Berry phases $\gamma_\text{B}$ of the RWA Hamiltonian. Within our reasoning, we expect that these phases are the FB phases of the Floquet states of the original Hamiltonian. This suggests that within RWA, the FB phases and the Berry phases are equal. Henceforth, we demonstrate that this is indeed the case.

We start from a Hamiltonian $H(\bm x, t)$ with explicit periodic time dependence and slow variation of the parameters $\bm x$.  We assume that by a $T$-periodic unitary matrix $U_0(t)$, the Hamiltonian can be transformed such that it becomes
\begin{align}
    \tilde H(\bm x,t)
    &{}= U_0^\dagger(t) H(\bm x,t)U_0(t) - iU_0^\dagger(t) \dot U_0(t)
    \\
    & \stackrel{\text{RWA}}{\longrightarrow}
    \bar H(\bm x) .
\end{align}
Technically, the second step is to replace the time-dependent interaction-picture Hamiltonian $\tilde H$ by its time average $\bar{H}$. This represents a suitable approximation when the magnitudes of the matrix elements of $\tilde H$ are much smaller than $\Omega$, which enables time-scale separation. Within this approximation, the solutions of the Schr\"odinger equation are of the form $|\psi(t)\rangle = e^{-iEt}|v\rangle$, where $|v\rangle$ and $E$ are eigenvector and eigenvalue of $\bar H$.  The Floquet state that corresponds to this solution reads
\begin{equation}
    |\phi(t)\rangle = U_0(t)|v\rangle ,
\end{equation}
with the quasienergy $q=E$ and the Floquet-Berry phase
\begin{align}
    \gamma_\text{FB}
    &{}= \oint d\bm x \cdot \langle v|U_0^\dagger(t) i\partial_{\bm x}U_0(t)|v\rangle
    \\
    &{}= \gamma_\text{B} .
\end{align}
This relation relies on two properties of the transformation $U_0(t)$. First, it must be $T$-periodic such that $U_0(t)|v\rangle$ is $T$-periodic as well, as is requested for a Floquet state.  Second, it must not depend on the parameters $\bm x$ such that $U_0$ commutes with the gradient $\partial_{\bm x}$ and cancels against $U_0^\dagger$.

The Hamiltonian of Leek \textit{et al.}~\cite{LeekS07} (following their notation, we interchange $\sigma_x$ and $\sigma_z$),
\begin{equation}
    H(t) = \frac{\Omega+\delta}{2}\sigma_z + A\sigma_x\cos(\Omega t+\theta),
\end{equation}
with $\delta,A \ll\Omega$ fulfills the conditions of an RWA treatment.  The unitary transformation with
\begin{equation}
    U_0(t) = \begin{pmatrix}
        e^{-i\Omega t} & 0 \\ 0 & 1
    \end{pmatrix}
\end{equation}
provides the RWA Hamiltonian
\begin{equation}
    \bar H = -\frac{\Omega}{2}\mathbb{1} + \frac{\delta}{2}\sigma_z
    + \frac{A}{2}(\sigma_x\cos\theta + \sigma_y\sin\theta) .
\end{equation}
Its Berry phase is given by half the solid angle enclosed by the trajectory $(A\cos\theta,A\sin\theta,\delta)$ for $\theta=0\ldots2\pi$ \cite{RestaJP00,XiaoRMP10}. This reveals that the protocol of Ref.~\cite{LeekS07} is a limiting case of the present one.

\section{Noise and dephasing}
In the protocol of Ref.~\cite{LeekS07}, the transverse Bloch vector is measured as a single complex number. For a given noise realization $\lambda$,
\begin{equation}
z_\lambda=\langle\sigma_x\rangle_\lambda+i\langle\sigma_y\rangle_\lambda=e^{i\phi_\lambda},
\quad \phi_\lambda=4\gamma(\lambda) ,
\label{eq:z}
\end{equation}
with $\gamma$ the geometric phase of a single loop. The analysis below holds for a generic geometric phase, irrespective of whether it is the AA phase or its adiabatic limit. For both, a noise-free measurement yields a pure final state that lies on the equator of the Bloch sphere, $|z|=1$. For a single static noise realization the evolution is unitary with a shifted parameter $\lambda$, so the same holds, $|z_\lambda|=1$, as long as the protocol remains adiabatic at $\lambda$. The experiment measures the ensemble average, $\bar{z}=\mathcal{V}e^{i\phi_\text{meas}}$, where the overline denotes the average over noise realizations. 

Any reduced modulus $|\bar{z}|$ corresponds to dephasing leading to visibility and phase
\begin{equation}
\begin{split}
\mathcal{V}={}& \big|\bar{z}\big|=\big|\overline{e^{i\phi_\lambda}}\big|\le1,
\\
\phi_\mathrm{meas} ={}& \arg\bar{z} = 4\gamma(\lambda_0) + \mathcal{O}(\sigma_\lambda^2),
\end{split}
\label{eq:contrast}
\end{equation}
respectively, where the former quantifies the coherence loss. Although dephasing generally has a dynamical and geometric contribution, the echo cancels the dynamical phase, and hence, its contribution to dephasing, for noise that is static over the protocol. Therefore, the main contribution to dephasing comes from the geometric phase $\gamma$ \cite{LeekS07}.
We adapt the dephasing model of Ref.~\cite{LeekS07} to our needs. This will highlight some differences between the two protocols and allow us to give a prediction of the dephasing of a possible experimental realization.
We address two pertinent questions: First, which noise frequencies couple to the geometric phase for a given loop frequency $\Omega/N_\text{loop}$ and, second, how strongly the phase of a single realization depends on the noise via the sensitivity $|\partial_\lambda\phi_\lambda|_{\lambda_0}=4|\partial_\lambda\gamma|$.

 \subsection{Noise sensitivity}
 \begin{figure}
  \centering
  \includegraphics[width=0.9\columnwidth]{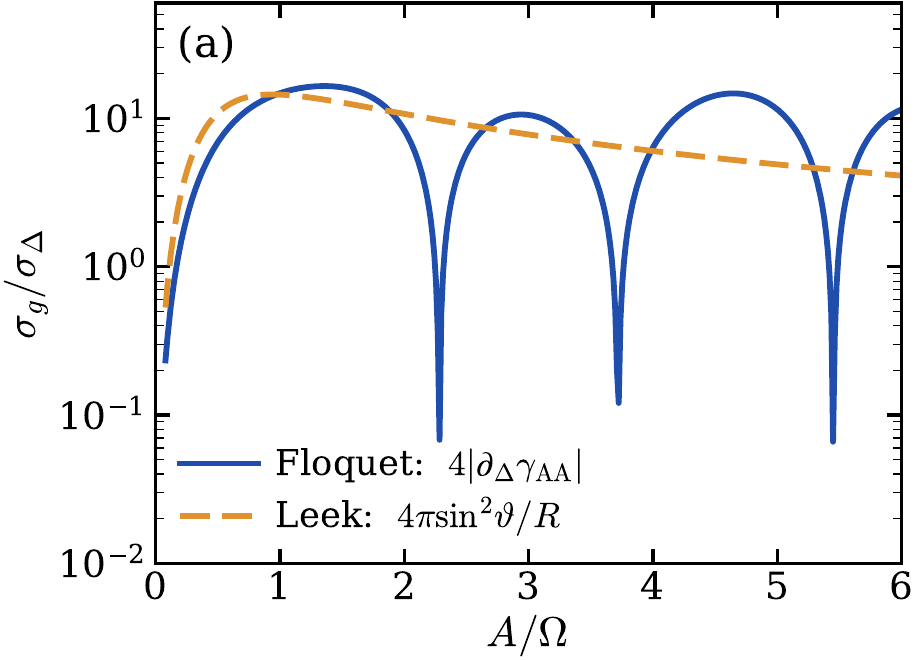}\\[2pt]
  \includegraphics[width=0.9\columnwidth]{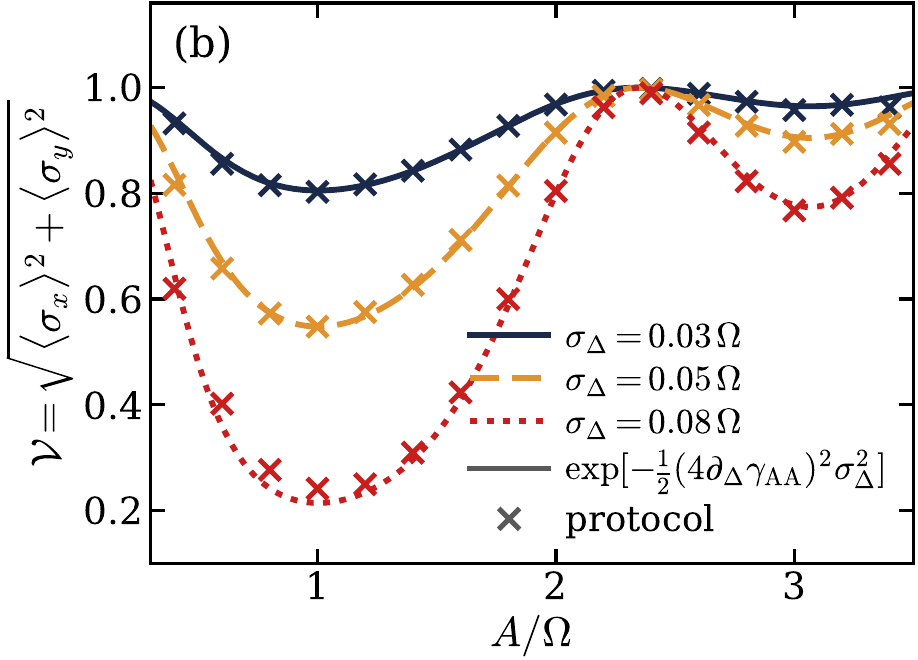}
\caption{Geometric dephasing under quasistatic Gaussian noise in the tunnel splitting, $\Delta\rightarrow\Delta+\delta\Delta$, with $\overline{\delta\Delta}=0$ and $\overline{\delta\Delta^2}=\sigma_\Delta^2$, as a function of the driving amplitude $A$ for $\Delta=\Omega/2$.
(a) Noise sensitivity of the measured phase after both forward and backward loops, $\sigma_g/\sigma_\Delta=\left|\partial_\Delta\phi_\lambda\right|$, in units of $\Omega^{-1}$. Solid: Floquet protocol, $4\left|\partial_\Delta\gammaAA\right|$ [Eq.~\eqref{eq:fl:dephasing}], for the Floquet mode $|\phi_-\rangle$ reached by the amplitude ramp from the ground state, at $\epsilon=\Omega/3$. The zeros correspond to the extrema of $\gammaAA(A)$, where $\phi_\lambda$ is insensitive to $\delta\Delta$ to first order.
Dashed: adiabatic Berry-phase protocol of Ref.~\cite{LeekS07} within the RWA, with detuning $\delta$ between the qubit and drive frequencies, path radius $\Omega_R$, $R=(\Omega_R^2+\delta^2)^{1/2}$, and cone angle $\cos\vartheta=\delta/R$. The curve uses $\Omega_R=A/2$, $\delta=\Omega/3$, and $\sigma_\omega=\sigma_\Delta$.
(b) Visibility $\mathcal{V}=(\langle\sigma_x\rangle^2+\langle\sigma_y\rangle^2)^{1/2}$ of the noise-averaged Bloch vector for $\epsilon=0.6\,\Omega$ and $\sigma_\Delta=0.03\,\Omega$ (solid), $0.05\,\Omega$ (dashed), and $0.08\,\Omega$ (dotted). Lines: Eq.~\eqref{eq:vis:noise}, $\mathcal{V}=\exp[-\frac{1}{2}(4\partial_\Delta\gammaAA)^2\sigma_\Delta^2]$. Crosses: simulations of the full protocol with $N_{\text{loop}}=40$ and $t_{\text{ramp}}=80\,T$; the Gaussian average is evaluated using 15-point Gauss--Hermite quadrature.}
\label{fig:sup:dephasing}
\end{figure}
Let us consider a Hamiltonian parameter carrying noise, $\lambda$ and a generic geometric phase built from a normalized state $|u(\theta)\rangle$ which is transported around the closed loop $\theta:0\rightarrow2\pi$, its geometric phase is defined as the holonomy of the Berry connection,
\begin{equation}
\gamma=\oint_0^{2\pi}\! A_\theta\,d\theta,\quad
A_\theta=i\,\langle u(\theta)|\partial_\theta u(\theta)\rangle\in\mathbb{R},
\label{eq:gamma-generic}
\end{equation}
where $A_\theta$ is real owing to normalization, $\langle u|u\rangle=1$. The change of the geometric phase with a parameter $\lambda$ can be studied by direct differentiation of Eq.~\eqref{eq:gamma-generic}. For that we will add and subtract the connection $A_\lambda$. This is done such that the result is expressed with respect to gauge invariant quantities as the direct differentiation of the connection $A_\theta$ is a gauge dependent quantity.
The manipulation of the first two terms yields
\begin{align}
\partial_\lambda A_\theta-\partial_\theta A_\lambda
&=i\big(\langle\partial_\lambda u|\partial_\theta u\rangle
      +\langle u|\partial_\theta\partial_\lambda u\rangle\big)\notag\\
&\quad -i\big(\langle\partial_\theta u|\partial_\lambda u\rangle
      +\langle u|\partial_\lambda\partial_\theta u\rangle\big)\notag\\
&=i\big(\langle\partial_\lambda u|\partial_\theta u\rangle-\langle\partial_\theta u|\partial_\lambda u\rangle\big)\notag\\
&=-2\,\mathrm{Im}\,\langle\partial_\lambda u|\partial_\theta u\rangle\;\equiv\;\mathcal{F}_{\lambda\theta},
\label{eq:curv}
\end{align}
where the mixed second derivatives cancel ($\partial_\theta\partial_\lambda u=\partial_\lambda\partial_\theta u$)
and $\mathcal{F}_{\lambda\theta}$ is the gauge invariant Berry curvature. Then,
\begin{equation}
\partial_\lambda\gamma=\oint \partial_\lambda A_\theta\,d\theta
=\underbrace{\oint\partial_\theta A_\lambda\,d\theta}_{=\,0}
+\oint \mathcal{F}_{\lambda\theta}\,d\theta,
\end{equation}
while the boundary term vanishes because $A_\lambda$ is single-valued across a closed loop. Finally, the change of the geometric phase as a function of the noise parameter $\lambda$ is given by
\begin{equation}
\partial_\lambda\gamma=\oint_0^{2\pi}\mathcal{F}_{\lambda\theta}(\theta)\,d\theta.
\label{eq:sensitivity}
\end{equation}
This implies that the sensitivity of the geometric phase to the noise is the loop integral of the Berry curvature. Therefore, noise only enters the geometric phase through the Berry curvature, i.e., through the change of the enclosed flux, which for the Berry phase of Leek \textit{et al.} is the enclosed solid angle.  This result is generic and valid for both Berry and AA phases, so we can treat both cases with the same expressions.

\paragraph{Berry phase, adiabatic \cite{LeekS07}:} The normalized state $|u(\theta)\rangle$ is the instantaneous eigenstate of the RWA field $\mathbf{R}=(\Omega_R \cos \theta,\Omega_R \sin\theta,\delta)$, swept slowly around a cone. Here, $\delta=\omega_a-\Omega$ is the detuning between the qubit transition frequency $\omega_a$ and the drive frequency, $\Omega_R$ is the radius of the circular path,  $R=|\mathbf{R}|=(\Omega_R^2+\delta^2)^{1/2}$, and $\vartheta$ is the cone angle, $\cos \vartheta=\delta/R$. A single loop yields the relative phase $\gamma_C=2\pi(1-\cos\vartheta)$, i.e., the enclosed angle. For fluctuations of $\omega_a$ with variance $\sigma_\omega^2$, Ref.~\cite{LeekS07} states the phase variance $\sigma_g^2=\sigma_\omega ^2 (2\pi\sin^2\vartheta/R)^2$; while $|\partial_\delta \gamma|=\pi \sin^2\vartheta/R$, which follows from Eq.~\eqref{eq:sensitivity} with $\lambda=\delta$. As the measured phase contains the relative phase of both eigenstates in both echo halves, the total sensitivity is $|\partial_\delta\phi|=4\pi\sin^2\vartheta/R$.

\paragraph{AA phase, non-adiabatic (this work):} $|u(\theta)\rangle$ is now the Floquet state $|\phi(\theta)\rangle$ which is an element of Sambe space. The sensitivity to noise can now be studied by differentiating the AA phase~\eqref{anandan_formula} and applying first-order Floquet perturbation theory to the Floquet operator $\mathcal{H}$:
\begin{equation}
\partial_\lambda\gamma_\alpha
=4\pi\,\mathrm{Re}\!\sum_{\beta\neq\alpha}
\frac{\langle\langle{\phi_\alpha}|i\partial_\theta |\phi_\beta\rangle\rangle  \langle\langle{\phi_\beta}|\partial_\lambda \mathcal{H}|\phi_\alpha\rangle\rangle}
{q_\alpha-q_\beta} .
\label{eq:sensitivityFloquet}
\end{equation}
The sum runs over all Floquet states including the replicas $|\phi_\alpha^{(\ell)}\rangle\rangle$. First-order perturbation theory requires the fluctuations to be small compared with the quasienergy gap, $\sigma_\lambda|\langle\langle \phi_\beta|\partial_\lambda\mathcal{H}|\phi_\alpha\rangle\rangle|\ll\Delta q$. The noise sensitivity diverges at resonances, where the quasienergy gap closes. We have already stated that the protocol is valid in regions where these resonances are avoided, so that our noise sensitivity derivation is still valid in experiment-friendly regions. In both cases, the sensitivity scales inversely with the relevant gap, $R$ and $\Delta q$, respectively. However, the quasienergy gap can become smaller than $R$, especially in regions close to CDT or multiphoton points, which enhances the sensitivity.

\subsection{Noise model}
The noise parameter fluctuates about its set point and can be parametrized as
\begin{equation}
    \lambda(t)=\lambda_0+\delta\lambda(t),
\end{equation}
with $\delta\lambda(t)$ a small, zero-mean, stationary, classical Gaussian fluctuation. The dephasing will be characterized by two properties of $\delta\lambda(t)$: how large it is, and how fast it moves.
\begin{itemize}
    \item \textbf{Size:} determined by the variance $\sigma_\lambda^2=\overline{\delta\lambda^2}$.
    \item \textbf{Speed:} the noise spectrum $S_\lambda(\omega)$, which tells how much of the variance sits at each frequency, i.e., how the noise power is distributed in frequency. Formally, the noise spectrum is defined as the Fourier transform of the autocorrelation function (the Wiener-Khinchin theorem~\cite{ClerkRMP10}) and its total integration yields the variance.
\end{itemize}
\begin{equation}
    \begin{gathered}
    S_\lambda(\omega)=\int_{-\infty}^\infty \overline{ \delta\lambda(t) \delta \lambda(t+\tau) } e^{-i\omega\tau}\, d\tau,\\
    \sigma_\lambda^2=\int\frac{d\omega}{2\pi}S_\lambda(\omega).
    \end{gathered}
    \label{eq:spectrum}
\end{equation}
For slow noise, $S_\lambda$ will mainly center around $\omega=0$, while a fast noise pushes $\omega$ to larger values. Therefore, dephasing strongly depends on the spectral density.
The noise model is rather generic, because the different noise sources merely require a proper choice of $\delta\lambda(t)$. 
Note that for the $1/f$ noise typical of superconducting qubits, the integral for $\sigma_\lambda^2$ diverges logarithmically and requires cutoffs.

However, we have not yet linked the noise model with the visibility, which is the quantity that will end up measuring the dephasing of our quantum system. By the sensitivity result~\eqref{eq:sensitivity}, a fluctuation acting at a point of the loop $\theta$ modifies the geometric phase through the local Berry curvature $\mathcal{F}_{\lambda\theta}(\theta)$. Taking this into account, the phase error is then the weighted average of the noise along the loop,
\begin{equation}
    \delta\phi_\lambda=4\oint_0^{2\pi} \mathcal{F}_{\lambda\theta}(\theta)\delta\lambda(\theta)d\theta,
    \label{eq:phase_error}
\end{equation}
which is linear in $\delta\lambda$. For Gaussian noise, the visibility~\eqref{eq:contrast} will depend only on the second moment, 
$\overline{e^{i\delta\phi_\lambda}}=e^{-\overline{\delta\phi_\lambda^2}/2}$
, and averaging~\eqref{eq:phase_error} with the noise spectrum~\eqref{eq:spectrum} yields
\begin{equation}
    \mathcal{V}=e^{-\sigma_g^2/2}, \quad \sigma_g^2=\int\frac{d\omega}{2\pi} S_\lambda(\omega)|\mathcal{F}(\omega)|^2,
    \label{noise}
\end{equation}
 where $\mathcal{F}(\omega)$ is the geometric filter, i.e., the loop average weighted by the phase that a frequency-$\omega$ noise fluctuation accumulates while the loop is traversed,
 \begin{equation}
     \mathcal{F}(\omega)=4\oint \mathcal{F}_{\lambda\theta}(\theta)e^{-i\omega \frac{N_\text{loop}\theta}{\Omega}}d\theta.
     \label{eq:filter}
 \end{equation}
 Equations~\eqref{eq:phase_error} and~\eqref{eq:filter} treat the two echo legs as sampling the same noise, which is exact for $\omega\tau\ll1$, with $\tau$ the separation between the legs. This is the regime valid for Eqs.~\eqref{eq:vis:noise}-\eqref{eq:fl:dephasing}.

\subsection{Dephasing during the protocol}

As we have seen, dephasing is controlled by where the spectral weight is located relative to the loop frequency $\omega_{\text{loop}}=\Omega/N_{\text{loop}}$. Two possible noises may play a role.
First, the quasistatic or slow noise ($\omega_{\text{noise}}\ll\omega_\text{loop}$). The noise is frozen over the whole echo sequence, $\delta\lambda(t)\simeq\delta\lambda$ for $t\in[0,T_\text{seq}]$. Then,
\begin{equation}
    \overline{z}=\overline{ e^{i4\gamma(\lambda_0+\delta \lambda)}}\simeq e^{i4\gamma(\lambda_0)}e^{-(4\partial_\lambda \gamma)^2\sigma_\lambda^2/2}.
\end{equation}
This gives the geometric dephasing law through the visibility
\begin{equation}
    \mathcal{V}=e^{-\sigma_g^2/2}, \quad \sigma_g=|\partial_\lambda \phi_\lambda| \sigma_\lambda=4|\partial_\lambda\gamma|\sigma_\lambda.
    \label{eq:vis:noise}
\end{equation}
It is the quasistatic limit of Eq.~\eqref{noise} and holds as long as the linear expansion is valid, $\sigma_\lambda|\partial_\lambda^2 \gamma|\ll |\partial\lambda \gamma|$; while close to the zeros of $\partial_\lambda \gamma$ the residual dephasing is of order $(\partial_\lambda^2\gamma~\sigma_\lambda^2)^2$.
The geometric dephasing is general and has the form $\exp(-\sigma_g^2/2)$ \cite{DeChiaraPRL03}. As we have seen, it is controlled by the loop frequency and not by the external drive frequency, so that our protocol obeys the dephasing law as for the experiment in the adiabatic limit \cite{LeekS07}. A difference may come on the sensitivity. By using the sensitivity for the Floquet states and specializing to fluctuating tunneling  $\lambda=\Delta$; the standard deviation for our protocol reads
\begin{equation}
    \sigma_g^\text{Floquet}=4\sigma_\Delta |\partial_\Delta\gammaAA|.
    \label{eq:fl:dephasing}
\end{equation}
Lastly, we should address the fast noise, when $\omega_{\text{noise}}\gg\omega_\text{loop}$. In this case, the noise oscillates many times inside the loop integral against the Berry curvature and averages to zero. Hence, fast noise does not dephase the geometric phase, but is averaged out. This holds for $\omega_\text{noise}\ll \Delta q$; noise at the transition frequencies $|q_\beta-q_\alpha+\ell\Omega|$ induces transitions between Floquet states, which is not captured by this dephasing model. This result, also noted by Leek \textit{et al.} holds for our protocol, so that our dephasing model just needs to focus on the aforementioned quasistatic noise.

Figure~\ref{fig:sup:dephasing} summarizes the study of the geometric dephasing under quasistatic Gaussian noise in the tunnel splitting for our model. First, Fig.~\ref{fig:sup:dephasing}(a) shows the behavior of the sensitivity for the adiabatic RWA case \cite{LeekS07} and our non-adiabatic Floquet protocol. 
Both are compared for the same observable $\phi_\lambda$ and are of the same order of magnitude, so that for comparable noise amplitude and sequence duration, the experimental realization of our protocol is comparable to that of Ref.~\cite{LeekS07}.
In addition, Floquet sensitivity reflects the zeros of the $\gammaAA$ phase of a strongly driven qubit, where the sensitivity is invariant to $\delta\Delta $ to first order.
On the other hand, Fig.~\ref{fig:sup:dephasing}(b) is an explicit confirmation of the geometric law of the dephasing. Simulated data of our protocol for $\sigma_\Delta=0.03\Omega,~0.05\Omega$ and $0.08\Omega$ is compared with the model Eq.~\eqref{eq:vis:noise}, showing excellent agreement for static noise, as assumed.

\end{document}